\documentclass[aps,prl,twocolumn,groupedaddress,showpacs]{revtex4-1}
\usepackage{graphicx}
\usepackage{dcolumn}
\usepackage{bm}
\usepackage{hyperref}
\usepackage{upgreek}

\begin{document}
\title{Coincidence of Superparamagnetism and perfect quantization in the Quantum Anomalous Hall state}

\author{S. Grauer}
\author{S. Schreyeck}
\author{M. Winnerlein}
\author{K. Brunner}
\author{C. Gould}
\author{L. W. Molenkamp}
\affiliation{Faculty for Physics and Astronomy and R\"ontgen Center for Complex Material Systems,
Universit\"at W\"urzburg, Am Hubland, D-97074, W\"urzburg, Germany}

\date{\today}

\begin{abstract}
Topological insulators doped with transition metals have recently been found to host a strong ferromagnetic state with perpendicular to plane anisotropy as well as support a quantum Hall state with edge channel transport, even in the absence of an external magnetic field. It remains unclear however why a robust magnetic state should emerge in materials of relatively low crystalline quality and dilute magnetic doping. Indeed, recent experiments suggest that the ferromagnetism exhibits at least some superparamagnetic character. We report on transport measurements in a sample that shows perfect quantum anomalous Hall quantization, while at the same time exhibits traits in its transport data which suggest inhomogeneities. We speculate that this may be evidence that the percolation path interpretation used to explain the transport during the magnetic reversal may actually have relevance over the entire field range.
\end{abstract}
\maketitle

The recent report on the experimental observation of a quantum anomalous Hall effect (QAHE) in Cr-doped (Bi,Sb)$_2$Te$_3$ \cite{Chang2013} generated significant interest in this material system for its potential as a magnetic topological insulator and as a test bed for the study of the Quantum Hall effect without the need for an external magnetic field \cite{Checkelsky2014,Kou2014,Bestwick2015,Chang2015,Chang2015b}. This original report showed that the anomalous Hall contribution \cite{Chang2013} appeared to saturate to a value of one conduction quantum as the sample was cooled to mK temperatures, but did not yet provide evidence that the transport takes place in edge states. \newline
In order to convincingly verify that the transport takes place in quantum Hall like edge states, non-local geometries are required. Such measurements were first reported in \cite{Kou2014}, albeit in configurations where the signals were very small, and where their interpretation required invoking some loss mechanism in the edge channels, and subsequently in \cite{Bestwick2015}, where convincing evidence of edge state transport was reported. This last paper also observed some unusual temperature and sweep rate related features in their data, which were at least in part interpreted as additional cooling through adiabatic demagnetization mechanisms. \newline
Shortly after the first reports on Cr-doped layers, it was discovered by the Moodera group \cite{Li2015,Chang2015,Chang2015b} that using V instead of Cr appears to lead to more reproducible samples with a more robust magnetic and quantum anomalous Hall state. Using this material system, the authors were able to reproduce both precise quantization of the Quantum Hall state \cite{Chang2015}, and unequivocal evidence of edge state transport \cite{Chang2015b}. \newline
While the described quantum anomalous Hall phenomenology is now well established, its microscopic origin remains much less clear. The proposed mechanism for the QAHE is the breaking of time reversal symmetry by a perpendicular to plane internal magnetic field which leads to the reversal of the band inversion of one of the two spin species in a ferromagnetic two dimensional topological insulator \cite{Chang2013,Yu2010}. The origin of the ferromagnetic state in the original paper \cite{Chang2013} was attributed to Van-Vleck ferromagnetism, which was first described by Bloembergen and Rowland \cite{Bloembergen1955}.\newline
It is unclear however why such a perfect magnetic state should come to occur in dilute magnetic semiconductors, nor how this interpretation is to be reconciled with SQUID visualization of the magnetic reversal in such layers, which show clear superparamagnetic behavior with magnetic clusters on the nanometer scale \cite{Lachman2015}. It is also unclear how such perfect transport behavior emerges from a material in which random dilute magnetic doping leads to a strongly locally disordered band gap \cite{Lee2015}. \newline
Finally it is surprising that such a state with a topological origin can be observed in this material with far from perfect crystal quality. Bi$_2$Se$_3$ and related compounds are well known to exhibit rotational twinning inherent to the crystal structure which can only be avoided for some specific substrate morphology \cite{Lachman2015,Li2010,Bansal2011,Wang2011,Tarakina2012,Guo2013,Schreyeck2013,Tarakina2014}. All the above may suggest that the percolation picture used for the description of transport during the reversal of the magnetic orientation \cite{Wang2014} may indeed be applicable for the quantized QAHE state as well. \newline   

In this paper we report on transport experiments of the QAHE state in a V-doped (Bi,Sb)$_{2}$Te$_{3}$ layer, and show that it exhibits some temperature and relaxation time related anomalous behavior. While a microscopic explanation for these anomalies has not yet been identified, we speculate that their understanding may lead to insight on the true nature of the magnetization in these samples. \newline
Our layer is grown by molecular beam epitaxy (MBE) on a hydrogen passivated Si(111) wafer, and capped in-situ with 10 nm amorphous Te to protect the topological insulator from the environment. The composition of the film is determined by comparison to electron dispersive X-ray spectroscopy on bulk reference samples grown under identical conditions. For the main layer used in this study, the composition is found to be V$_{0.11}$Bi$_{0.445}$Sb$_{1.40}$Te$_{3.045}$, while the thickness of the investigated film is determined to be 10 nm by X-ray reflection measurements. XRD observation on these samples confirm that these layers are indeed heavily rotational twinned, with roughly equal representation of each twin,  and thus nowhere near homogeneous single crystals.\newline
The grown film is patterned into a gated Hall bar geometry using standard optical lithography, with a layer comprised of 20 nm of AlOx and 1 nm of HfOx grown conformally by atomic layer deposition providing insulation for the top gate. An image of the resulting sample is shown in fig. 1. The two investigated Hall bars, big and (small) have a width of 200 (10)  $\upmu$m, and a separation of 600 (30) $\upmu$m between adjacent contacts. The numbered contacts in fig. 1 belong to the large Hall bar.\newline
After the sample was cooled, in a dilution refrigerator equipped with a high field magnet, Hall and longitudinal resistance measurements were performed to determine the gate voltage range where the topological insulator bulk is in its insulating state, and where, as a result, one observes the maximum value in the Hall signal, and the minimum longitudinal resistance. All data presented in this paper is taken at the gate voltage in the center of this range. \newline 
Figure 1 shows results of the basic experiment. The sample is cooled to a base temperature of nominally 15 mK, and its Hall and Magnetoresitive behavior is measured using standard high precision low frequency ac techniques (2 to 14 Hz). In the top panel, a traditional configuration is used, where a current is passed from contacts 1 to 4, and determined by measuring the voltage drop across a calibrated resistor in series with the device. The longitudinal ($\rho _{xx}$) and transversal ($\rho _{xy}$) resistances are then determined from measuring the voltages between 2 and 3, or between 3 and 5, respectively. \newline
\begin{figure}
\includegraphics[width=\columnwidth]{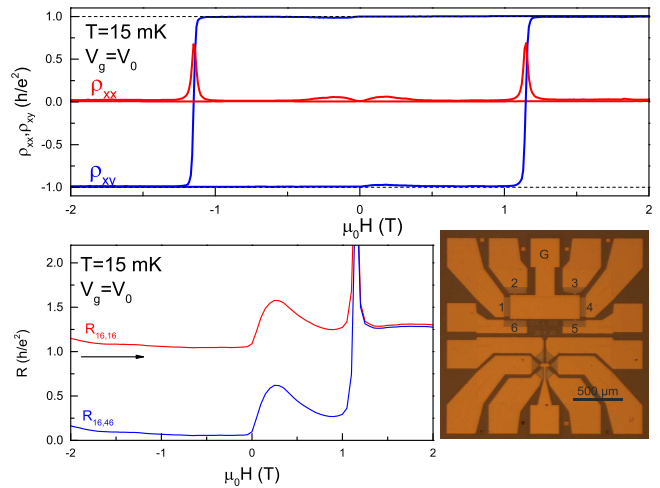}
\caption{\label{fig:fig1} Top: $\rho _{xx}$ and $\rho _{xy}$ as a function of magnetic field. Bottom: Non-local measurement evidencing the edge state transport. For both curves, current flows from contact 1 to the ground at 6 (see picture on the right). For the red curve the voltage is also detected at 1 and 6, whereas for the blue curve it is measured between 4 and 6. On the right is an optical photograph of the device with the numbering of the leads indicated for the large Hall bar. G denotes the gate contact. }
\end{figure} 
Ignoring for the moment the feature which appears shortly after the external applied field $H = 0$, the data looks similar to that published in Cr-doped  \cite{Chang2013,Checkelsky2014,Kou2014,Bestwick2015} and in V-doped samples \cite{Chang2015,Chang2015b}. The as measured value of $\rho _{xx}$ and $\rho _{xy}$ at fields just before this feature are 0.005 and 0.996 respectively, in units of the von Klitzing constant R$_{\text{K}}$. This small deviation from the von Klitzing constant results from current leakage through the impedance of the instrument and cryostat leads to the voltage probes. Since the capacitance of the various components is not accurately known, this contribution cannot be corrected for exactly, but an order of magnitude estimate is in agreement with the measured deviation for the literature value. \newline
In the lower part of the panel we show a non-local measurement to confirm that the transport takes place through edge states. $R_{ab,cd}$ is simply the voltage between $c$ and $d$ divided by the current flowing from $a$ to $b$, so for both the red and blue curves, the current is passed between 1 and the circuit ground at 6. In the edge state picture, the voltage injected into contact 1 will be transported around the edge of the device, clockwise or counter clockwise depending on the direction of the internal magnetic field, until it can be drained into the ground. For the red curve, which measured the voltage also between contacts 1 and 6, the measured value is near R$_{\text{K}}$ except for during the reversal of the magnetization, where the bulk becomes conducting and normal bulk transport is observed \cite{Chang2015b} and at the feature just after $H=0$. The small deviations from quantization at higher fields come from the contact resistances. \newline
In the case of the blue curve, the voltage is measured between 4 and 6. In the case of negative fields, as the current is flowing counter clockwise, directly from 1 to 6, no current flows between 4 and 6, and thus the resulting resistance is near 0. After the magnetization reverses at slightly above 1 T, the current begins to flow clockwise, and contact 4 is at the same potential as contact 1, leading to a measured resistance of one conductance quantum. Again the deviation from perfect quantization (and a perfect zero) results from imperfect contacts, as the finite resistance of the lead connected to contact 6 causes a series resistance and thus a voltage drop between the potential of the edge state to be measured and the measuring instrument. \newline
While the majority of the results in fig. 1 are in  agreement with previous works on both Cr- and V-doped samples, all curves in the figure show a resistance feature after $H=0$, which has not been previously reported on. It should be noted however that a similar feature is clearly visible for $\rho _{xx}$ in fig. 1b of \cite{Bestwick2015}, and indeed can also be seen for $\rho _{xy}$ if one knows to look for it. We now turn our attention to studying this feature in more detail. \newline
In the top part of the fig. 2 we see a zoomed in version of the $\rho _{xy}$ feature, and show that it is hysteretic in nature in that it is observed immediately upon crossing $H=0$ from either directions. Such a feature is inconsistent with the picture of a perfect macro-spin like ferromagnet, because in such a material, where the internal field $B = H + \mu M$, there is no particular feature in $B$ at $H = 0$. We therefore interpret this as further evidence that the magnetic state is more complex and quite possibly is of some superparamagnetic nature. The middle part of the figure shows curves taken at different temperatures showing that this feature, which is still clearly visible in the 530 mK curve, survives to temperatures well beyond those for which the anomalous Hall effect is quantized, but dies by 1 K, well before the macroscopic Curie temperature of $T_C=26\text{ K}$ (as determined by SQUID) is reached. \newline
\begin{figure}[!ht]
\includegraphics[width=\columnwidth]{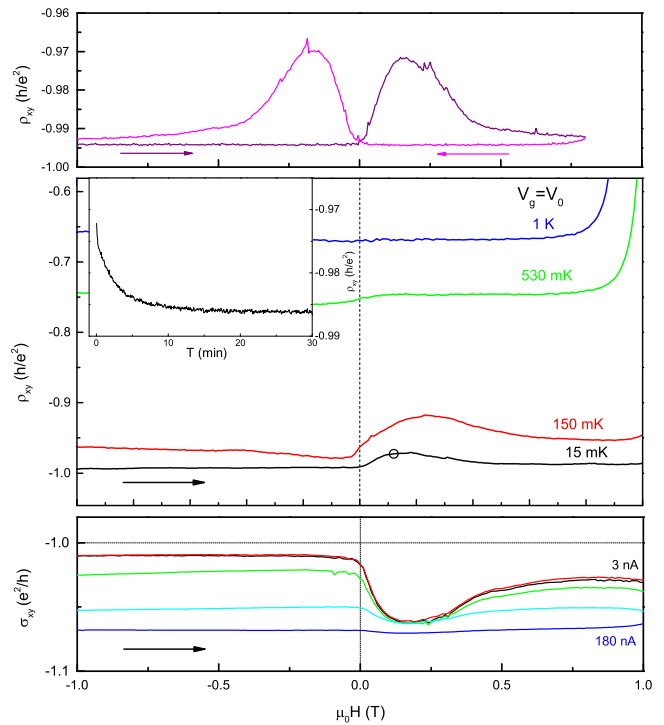}
\caption{\label{fig:fig2} Measurements focusing on the feature near $H = 0$. Top: A low temperature low current measurement showing the hysteretic nature of the feature. Middle: Temperature dependence of the feature, and in the inset, time dependence after the field sweep has been paused, for the conditions indicated by the black circle on the 15 mK curve of the main figure. Bottom panel: Current dependence of the feature.}
\end{figure} 
In the bottom panel of the figure we show measurements of the same feature observed on a second sample (with about 6\% less Sb). In this case, $\sigma_{xy}$ is measured for various current levels ranging from 3 to 180 nA. We observe that large currents can suppress the feature, perhaps hinting at a spin torque contribution from the flowing current stabilizing the magnetic domains.\newline
Further evidence that this feature may relate to superparamagnetism or an akin magnetic viscosity is provided in the inset of the figure. Here, the 15 mK measurement of the main figure is repeated, sweeping the field up to the black circle. At this point, the field is held constant and the Hall resistance continues to be monitored. The feature relaxes partially away over the time scale of several minutes, but it should be noted that it relaxes back only to -0.986 R$_{\text{K}}$, and, on the observation timescale, not all the way to the baseline, which for this measurement configuration has a value of -0.993 R$_{\text{K}}$. \newline
While the low field feature may be the most prominent indication that there is something complex going on in the magnetic state, it is by no means the only one. In fig. 3 we highlight a region of field where the resistance exhibits spontaneous jumps. This noise is only observed in a region of external magnetic fields after the magnetic reversal, and up to a field corresponding to about trice the coercive field. The region where this noise is visible is reproducible when repeating the measurement, as seen from the 3 successive measurements plotted in various colors in the figure. The exact noise pattern however changes randomly from curve to curve. The right hand of the figure highlights that the noise is only visible after the switching event, which is to say that when sweeping the field up from negative field towards zero we do not observe it, but we do see it in the same field range when coming from positive fields, and thus switching the bulk magnetization. This too suggests some superparamagnetic like switching behavior, as it may indicate that shortly after the magnetic reversal, some domains are not stable and can spontaneously reverse their magnetization, causing domain walls to move around and generate stochastic jumps in the Hall resistance. Only when a sufficiently large H field is applied are all domains passivated, and a stable magnetic state is reached. This interpretation is further supported by the data shown in the bottom of fig. 3, which was measured on the second sample. For these measurements the magnetic state of the device was prepared by sweeping the external magnetic field from positive values to the specified starting field. The field was then swept back to positive values and we observe that the density of spikes in the curve depends strongly on the amplitude of the preparation field. The top curve is prepared, such that the magnetization of the sample is barely reversed, as the coercive field is about 1.1 T at base temperature, and no spikes are seen. If the absolute value of the preparation field is further increased, making more domains switch on the down-sweep, the spikes return. Interestingly, the density of peaks increases as the preparation field is made more and more negative. This suggests that preparing the field by sweeping it to say, 2.6 T, well beyond its coercive field, does not saturate the magnetic state, as it would in the case of a macrospin ferromagnet. Instead, the data implies that much larger fields are required to reverse all the magnetic domains, which then switch in the form of the spikes when the field is reversed in the final measurement. Consistent with this interpretation is that the reason these spikes are visible in these measurements and not in the data of fig. 1 has to do with a detail of the data acquisition. In fig. 1, the data was collected by allowing the external magnetic field to stabilize for 3 seconds at each field value before collecting the data, whereas in fig. 3 the measurements were done without settling the field. We further note that in the other region of magnetic instability, which is to say on the zero field features, similar, albeit of much lower amplitude, jumps in the $\rho _{xy}$ are observed. \newline
\begin{figure}[!ht]
\includegraphics[width=\columnwidth]{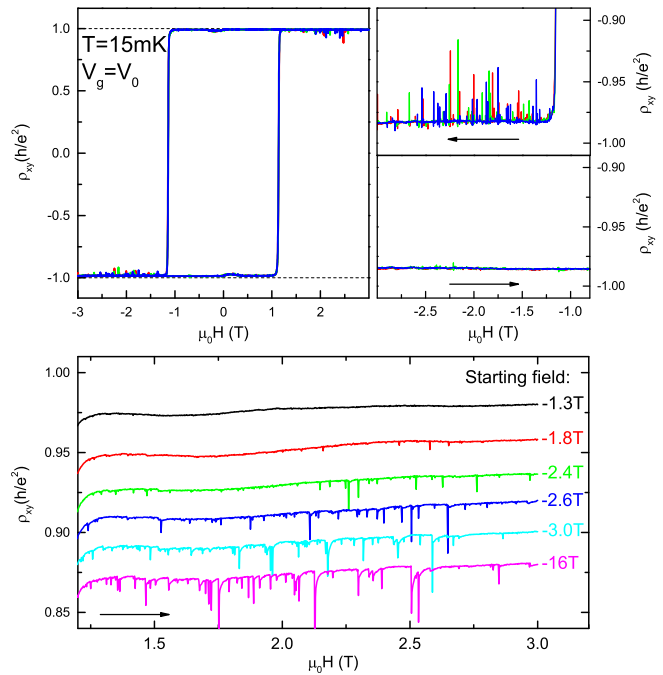}
\caption{\label{fig:fig3} $\rho _{xy}$ measured at low current, showing a series of spikes in resistance which occur at fields after the magnetic switching event under proper sweep conditions (see text for details). The right panels show that the spikes only occur in one field sweep direction as given by the arrows. Bottom panel: measurement taken on the second sample showing that the density of spikes depends upon how strongly the preparation state is saturated. (Curves offset vertically for clarity.)}
\end{figure} 
And lastly, one further peculiarity is observed in the temperature dependence of the anomalous Hall effect. The cooling curve in fig. 4 monitors the Hall resistance as the fridge is cooled from 1 K down to base. We see an already fairly large anomalous Hall effect at 1 K, which grows to its quantized value around 50 mK. This curve looks qualitatively as one might expect. The warming curve is a different matter altogether.  Shortly after beginning to warm the fridge, and at temperatures much lower than 50 mK, the Hall resistance drops considerably. More peculiar still is the effect of changing the warming rate of the fridge. Here was observed that such changes are often accompanied by strong changes in the Hall resistance. Great care was taken to verify that there is no short in the wiring, or defect in the heater power supply, and that the only influence on the sample is indeed the warming rate. The correspondence between changes in warming rates and jumps in the Hall resistance is not one-to-one perfect, but it is much too high to be a coincidence, and again suggest some stochasticly sensitive magnetic state. \newline
\begin{figure}[!ht]
\includegraphics[width=\columnwidth]{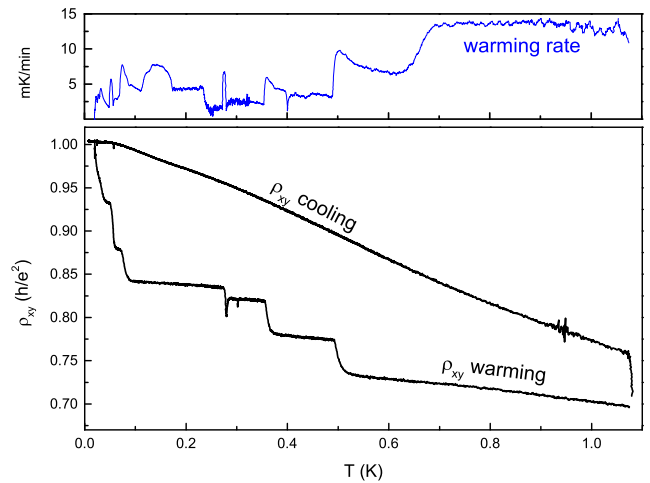}
\caption{\label{fig:fig4}  $\rho _{xy}$ as a function of temperature, highlighting the peculiar behavior that the resistance appears to at least partly depend on the rate of change of temperature. The top panel shows the time derivative of the mixing chamber temperature during warming.}
\end{figure} 
In summary, we have seen in V-doped (Bi,Sb)$_{2}$Te$_{3}$ which exhibits perfect QAHE quantization and clear edge state transport a number of anomalies in the transport behavior, which all point to instabilities in the magnetic state of the sample. This confirms the magnetic imaging work of \cite{Lachman2015} and strongly suggests that a macrospin type description of the magnetization is inappropriate for such samples. This may indeed indicate that a percolation picture of the transport phenomenology in these materials is not only appropriate for description of the transport at the transition between Hall plateaus \cite{Wang2014}, but that similar considerations may govern the entire field range. While we do not at present have a model for what type of superparamagnetic or magnetically viscous contributions might cause this phenomenology, we hope that publication of this data will stimulate modeling and theoretical work in this area and may lead to a better understanding of the magnetism in these interesting materials. \newline

\begin{acknowledgments}
We gratefully acknowledge the financial support of the EU ERC-AG Program (project 3-TOP) and the DFG through SFB 1170 ``ToCoTronics'' and the Leibniz Program, as well as useful discussions with J. Moodera, C.-Z. Chang, and C. Br\"une on device growth, discussions with A. MacDonald and N. Nagaosa on transport physics, and experimental assistance from D. Mahler and S. Rosenberger.
\end{acknowledgments}
\bibliography{RefsQAHE}

\begin{thebibliography}{19}%
\makeatletter
\providecommand \@ifxundefined [1]{%
 \@ifx{#1\undefined}
}%
\providecommand \@ifnum [1]{%
 \ifnum #1\expandafter \@firstoftwo
 \else \expandafter \@secondoftwo
 \fi
}%
\providecommand \@ifx [1]{%
 \ifx #1\expandafter \@firstoftwo
 \else \expandafter \@secondoftwo
 \fi
}%
\providecommand \natexlab [1]{#1}%
\providecommand \enquote  [1]{``#1''}%
\providecommand \bibnamefont  [1]{#1}%
\providecommand \bibfnamefont [1]{#1}%
\providecommand \citenamefont [1]{#1}%
\providecommand \href@noop [0]{\@secondoftwo}%
\providecommand \href [0]{\begingroup \@sanitize@url \@href}%
\providecommand \@href[1]{\@@startlink{#1}\@@href}%
\providecommand \@@href[1]{\endgroup#1\@@endlink}%
\providecommand \@sanitize@url [0]{\catcode `\\12\catcode `\$12\catcode
  `\&12\catcode `\#12\catcode `\^12\catcode `\_12\catcode `\%12\relax}%
\providecommand \@@startlink[1]{}%
\providecommand \@@endlink[0]{}%
\providecommand \url  [0]{\begingroup\@sanitize@url \@url }%
\providecommand \@url [1]{\endgroup\@href {#1}{\urlprefix }}%
\providecommand \urlprefix  [0]{URL }%
\providecommand \Eprint [0]{\href }%
\providecommand \doibase [0]{http://dx.doi.org/}%
\providecommand \selectlanguage [0]{\@gobble}%
\providecommand \bibinfo  [0]{\@secondoftwo}%
\providecommand \bibfield  [0]{\@secondoftwo}%
\providecommand \translation [1]{[#1]}%
\providecommand \BibitemOpen [0]{}%
\providecommand \bibitemStop [0]{}%
\providecommand \bibitemNoStop [0]{.\EOS\space}%
\providecommand \EOS [0]{\spacefactor3000\relax}%
\providecommand \BibitemShut  [1]{\csname bibitem#1\endcsname}%
\let\auto@bib@innerbib\@empty
\bibitem [{\citenamefont {Chang}\ \emph {et~al.}(2013)\citenamefont {Chang},
  \citenamefont {Zhang}, \citenamefont {Feng}, \citenamefont {Shen},
  \citenamefont {Zhang}, \citenamefont {Guo}, \citenamefont {Li}, \citenamefont
  {Ou}, \citenamefont {Wei}, \citenamefont {Wang}, \citenamefont {Ji},
  \citenamefont {Feng}, \citenamefont {Ji}, \citenamefont {Chen}, \citenamefont
  {Jia}, \citenamefont {Dai}, \citenamefont {Fang}, \citenamefont {Zhang},
  \citenamefont {He}, \citenamefont {Wang}, \citenamefont {Lu}, \citenamefont
  {Ma},\ and\ \citenamefont {Xue}}]{Chang2013}%
  \BibitemOpen
  \bibfield  {author} {\bibinfo {author} {\bibfnamefont {C.-Z.}\ \bibnamefont
  {Chang}}, \bibinfo {author} {\bibfnamefont {J.}~\bibnamefont {Zhang}},
  \bibinfo {author} {\bibfnamefont {X.}~\bibnamefont {Feng}}, \bibinfo {author}
  {\bibfnamefont {J.}~\bibnamefont {Shen}}, \bibinfo {author} {\bibfnamefont
  {Z.}~\bibnamefont {Zhang}}, \bibinfo {author} {\bibfnamefont
  {M.}~\bibnamefont {Guo}}, \bibinfo {author} {\bibfnamefont {K.}~\bibnamefont
  {Li}}, \bibinfo {author} {\bibfnamefont {Y.}~\bibnamefont {Ou}}, \bibinfo
  {author} {\bibfnamefont {P.}~\bibnamefont {Wei}}, \bibinfo {author}
  {\bibfnamefont {L.-L.}\ \bibnamefont {Wang}}, \bibinfo {author}
  {\bibfnamefont {Z.-Q.}\ \bibnamefont {Ji}}, \bibinfo {author} {\bibfnamefont
  {Y.}~\bibnamefont {Feng}}, \bibinfo {author} {\bibfnamefont {S.}~\bibnamefont
  {Ji}}, \bibinfo {author} {\bibfnamefont {X.}~\bibnamefont {Chen}}, \bibinfo
  {author} {\bibfnamefont {J.}~\bibnamefont {Jia}}, \bibinfo {author}
  {\bibfnamefont {X.}~\bibnamefont {Dai}}, \bibinfo {author} {\bibfnamefont
  {Z.}~\bibnamefont {Fang}}, \bibinfo {author} {\bibfnamefont {S.-C.}\
  \bibnamefont {Zhang}}, \bibinfo {author} {\bibfnamefont {K.}~\bibnamefont
  {He}}, \bibinfo {author} {\bibfnamefont {Y.}~\bibnamefont {Wang}}, \bibinfo
  {author} {\bibfnamefont {L.}~\bibnamefont {Lu}}, \bibinfo {author}
  {\bibfnamefont {X.-C.}\ \bibnamefont {Ma}}, \ and\ \bibinfo {author}
  {\bibfnamefont {Q.-K.}\ \bibnamefont {Xue}},\ }\href {\doibase
  10.1126/science.1234414} {\bibfield  {journal} {\bibinfo  {journal}
  {Science}\ }\textbf {\bibinfo {volume} {340}},\ \bibinfo {pages} {167}
  (\bibinfo {year} {2013})}\BibitemShut {NoStop}%
\bibitem [{\citenamefont {Checkelsky}\ \emph {et~al.}(2014)\citenamefont
  {Checkelsky}, \citenamefont {Yoshimi}, \citenamefont {Tsukazaki},
  \citenamefont {Takahashi}, \citenamefont {Kozuka}, \citenamefont {Falson},
  \citenamefont {Kawasaki},\ and\ \citenamefont {Tokura}}]{Checkelsky2014}%
  \BibitemOpen
  \bibfield  {author} {\bibinfo {author} {\bibfnamefont {J.~G.}\ \bibnamefont
  {Checkelsky}}, \bibinfo {author} {\bibfnamefont {R.}~\bibnamefont {Yoshimi}},
  \bibinfo {author} {\bibfnamefont {A.}~\bibnamefont {Tsukazaki}}, \bibinfo
  {author} {\bibfnamefont {K.~S.}\ \bibnamefont {Takahashi}}, \bibinfo {author}
  {\bibfnamefont {Y.}~\bibnamefont {Kozuka}}, \bibinfo {author} {\bibfnamefont
  {J.}~\bibnamefont {Falson}}, \bibinfo {author} {\bibfnamefont
  {M.}~\bibnamefont {Kawasaki}}, \ and\ \bibinfo {author} {\bibfnamefont
  {Y.}~\bibnamefont {Tokura}},\ }\href {\doibase 10.1038/nphys3053
  http://www.nature.com/nphys/journal/v10/n10/abs/nphys3053.html#supplementary-information}
  {\bibfield  {journal} {\bibinfo  {journal} {Nat Phys}\ }\textbf {\bibinfo
  {volume} {10}},\ \bibinfo {pages} {731} (\bibinfo {year} {2014})}\BibitemShut
  {NoStop}%
\bibitem [{\citenamefont {Kou}\ \emph {et~al.}(2014)\citenamefont {Kou},
  \citenamefont {Guo}, \citenamefont {Fan}, \citenamefont {Pan}, \citenamefont
  {Lang}, \citenamefont {Jiang}, \citenamefont {Shao}, \citenamefont {Nie},
  \citenamefont {Murata}, \citenamefont {Tang}, \citenamefont {Wang},
  \citenamefont {He}, \citenamefont {Lee}, \citenamefont {Lee},\ and\
  \citenamefont {Wang}}]{Kou2014}%
  \BibitemOpen
  \bibfield  {author} {\bibinfo {author} {\bibfnamefont {X.}~\bibnamefont
  {Kou}}, \bibinfo {author} {\bibfnamefont {S.-T.}\ \bibnamefont {Guo}},
  \bibinfo {author} {\bibfnamefont {Y.}~\bibnamefont {Fan}}, \bibinfo {author}
  {\bibfnamefont {L.}~\bibnamefont {Pan}}, \bibinfo {author} {\bibfnamefont
  {M.}~\bibnamefont {Lang}}, \bibinfo {author} {\bibfnamefont {Y.}~\bibnamefont
  {Jiang}}, \bibinfo {author} {\bibfnamefont {Q.}~\bibnamefont {Shao}},
  \bibinfo {author} {\bibfnamefont {T.}~\bibnamefont {Nie}}, \bibinfo {author}
  {\bibfnamefont {K.}~\bibnamefont {Murata}}, \bibinfo {author} {\bibfnamefont
  {J.}~\bibnamefont {Tang}}, \bibinfo {author} {\bibfnamefont {Y.}~\bibnamefont
  {Wang}}, \bibinfo {author} {\bibfnamefont {L.}~\bibnamefont {He}}, \bibinfo
  {author} {\bibfnamefont {T.-K.}\ \bibnamefont {Lee}}, \bibinfo {author}
  {\bibfnamefont {W.-L.}\ \bibnamefont {Lee}}, \ and\ \bibinfo {author}
  {\bibfnamefont {K.~L.}\ \bibnamefont {Wang}},\ }\href
  {http://link.aps.org/doi/10.1103/PhysRevLett.113.137201} {\bibfield
  {journal} {\bibinfo  {journal} {Physical Review Letters}\ }\textbf {\bibinfo
  {volume} {113}},\ \bibinfo {pages} {137201} (\bibinfo {year}
  {2014})}\BibitemShut {NoStop}%
\bibitem [{\citenamefont {Bestwick}\ \emph {et~al.}(2015)\citenamefont
  {Bestwick}, \citenamefont {Fox}, \citenamefont {Kou}, \citenamefont {Pan},
  \citenamefont {Wang},\ and\ \citenamefont {Goldhaber-Gordon}}]{Bestwick2015}%
  \BibitemOpen
  \bibfield  {author} {\bibinfo {author} {\bibfnamefont {A.~J.}\ \bibnamefont
  {Bestwick}}, \bibinfo {author} {\bibfnamefont {E.~J.}\ \bibnamefont {Fox}},
  \bibinfo {author} {\bibfnamefont {X.}~\bibnamefont {Kou}}, \bibinfo {author}
  {\bibfnamefont {L.}~\bibnamefont {Pan}}, \bibinfo {author} {\bibfnamefont
  {K.~L.}\ \bibnamefont {Wang}}, \ and\ \bibinfo {author} {\bibfnamefont
  {D.}~\bibnamefont {Goldhaber-Gordon}},\ }\href
  {http://link.aps.org/doi/10.1103/PhysRevLett.114.187201} {\bibfield
  {journal} {\bibinfo  {journal} {Physical Review Letters}\ }\textbf {\bibinfo
  {volume} {114}},\ \bibinfo {pages} {187201} (\bibinfo {year}
  {2015})}\BibitemShut {NoStop}%
\bibitem [{\citenamefont {Chang}\ \emph {et~al.}(2015)\citenamefont {Chang},
  \citenamefont {Zhao}, \citenamefont {Kim}, \citenamefont {Zhang},
  \citenamefont {Assaf}, \citenamefont {Heiman}, \citenamefont {Zhang},
  \citenamefont {Liu}, \citenamefont {Chan},\ and\ \citenamefont
  {Moodera}}]{Chang2015}%
  \BibitemOpen
  \bibfield  {author} {\bibinfo {author} {\bibfnamefont {C.-Z.}\ \bibnamefont
  {Chang}}, \bibinfo {author} {\bibfnamefont {W.}~\bibnamefont {Zhao}},
  \bibinfo {author} {\bibfnamefont {D.~Y.}\ \bibnamefont {Kim}}, \bibinfo
  {author} {\bibfnamefont {H.}~\bibnamefont {Zhang}}, \bibinfo {author}
  {\bibfnamefont {B.~A.}\ \bibnamefont {Assaf}}, \bibinfo {author}
  {\bibfnamefont {D.}~\bibnamefont {Heiman}}, \bibinfo {author} {\bibfnamefont
  {S.-C.}\ \bibnamefont {Zhang}}, \bibinfo {author} {\bibfnamefont
  {C.}~\bibnamefont {Liu}}, \bibinfo {author} {\bibfnamefont {M.~H.~W.}\
  \bibnamefont {Chan}}, \ and\ \bibinfo {author} {\bibfnamefont {J.~S.}\
  \bibnamefont {Moodera}},\ }\href {\doibase 10.1038/nmat4204
  http://www.nature.com/nmat/journal/v14/n5/abs/nmat4204.html#supplementary-information}
  {\bibfield  {journal} {\bibinfo  {journal} {Nat Mater}\ }\textbf {\bibinfo
  {volume} {14}},\ \bibinfo {pages} {473} (\bibinfo {year} {2015})}\BibitemShut
  {NoStop}%
\bibitem [{\citenamefont {{Chang}}\ \emph {et~al.}(2015)\citenamefont
  {{Chang}}, \citenamefont {{Zhao}}, \citenamefont {{Kim}}, \citenamefont
  {{Wei}}, \citenamefont {{Jain}}, \citenamefont {{Liu}}, \citenamefont
  {{Chan}},\ and\ \citenamefont {{Moodera}}}]{Chang2015b}%
  \BibitemOpen
  \bibfield  {author} {\bibinfo {author} {\bibfnamefont {C.-Z.}\ \bibnamefont
  {{Chang}}}, \bibinfo {author} {\bibfnamefont {W.}~\bibnamefont {{Zhao}}},
  \bibinfo {author} {\bibfnamefont {D.~Y.}\ \bibnamefont {{Kim}}}, \bibinfo
  {author} {\bibfnamefont {P.}~\bibnamefont {{Wei}}}, \bibinfo {author}
  {\bibfnamefont {J.~K.}\ \bibnamefont {{Jain}}}, \bibinfo {author}
  {\bibfnamefont {C.}~\bibnamefont {{Liu}}}, \bibinfo {author} {\bibfnamefont
  {M.~H.~W.}\ \bibnamefont {{Chan}}}, \ and\ \bibinfo {author} {\bibfnamefont
  {J.~S.}\ \bibnamefont {{Moodera}}},\ }\href@noop {} {\bibfield  {journal}
  {\bibinfo  {journal} {arXiv:1505.01896}\ } (\bibinfo {year}
  {2015})}\BibitemShut {NoStop}%
\bibitem [{\citenamefont {Li}\ \emph {et~al.}(2015)\citenamefont {Li},
  \citenamefont {Chang}, \citenamefont {Wu}, \citenamefont {Tao}, \citenamefont
  {Zhao}, \citenamefont {Chan}, \citenamefont {Moodera}, \citenamefont {Li},\
  and\ \citenamefont {Zhu}}]{Li2015}%
  \BibitemOpen
  \bibfield  {author} {\bibinfo {author} {\bibfnamefont {M.}~\bibnamefont
  {Li}}, \bibinfo {author} {\bibfnamefont {C.-Z.}\ \bibnamefont {Chang}},
  \bibinfo {author} {\bibfnamefont {L.}~\bibnamefont {Wu}}, \bibinfo {author}
  {\bibfnamefont {J.}~\bibnamefont {Tao}}, \bibinfo {author} {\bibfnamefont
  {W.}~\bibnamefont {Zhao}}, \bibinfo {author} {\bibfnamefont {M.~H.~W.}\
  \bibnamefont {Chan}}, \bibinfo {author} {\bibfnamefont {J.~S.}\ \bibnamefont
  {Moodera}}, \bibinfo {author} {\bibfnamefont {J.}~\bibnamefont {Li}}, \ and\
  \bibinfo {author} {\bibfnamefont {Y.}~\bibnamefont {Zhu}},\ }\href
  {http://link.aps.org/doi/10.1103/PhysRevLett.114.146802} {\bibfield
  {journal} {\bibinfo  {journal} {Physical Review Letters}\ }\textbf {\bibinfo
  {volume} {114}},\ \bibinfo {pages} {146802} (\bibinfo {year}
  {2015})}\BibitemShut {NoStop}%
\bibitem [{\citenamefont {Yu}\ \emph {et~al.}(2010)\citenamefont {Yu},
  \citenamefont {Zhang}, \citenamefont {Zhang}, \citenamefont {Zhang},
  \citenamefont {Dai},\ and\ \citenamefont {Fang}}]{Yu2010}%
  \BibitemOpen
  \bibfield  {author} {\bibinfo {author} {\bibfnamefont {R.}~\bibnamefont
  {Yu}}, \bibinfo {author} {\bibfnamefont {W.}~\bibnamefont {Zhang}}, \bibinfo
  {author} {\bibfnamefont {H.-J.}\ \bibnamefont {Zhang}}, \bibinfo {author}
  {\bibfnamefont {S.-C.}\ \bibnamefont {Zhang}}, \bibinfo {author}
  {\bibfnamefont {X.}~\bibnamefont {Dai}}, \ and\ \bibinfo {author}
  {\bibfnamefont {Z.}~\bibnamefont {Fang}},\ }\href {\doibase
  10.1126/science.1187485} {\bibfield  {journal} {\bibinfo  {journal}
  {Science}\ }\textbf {\bibinfo {volume} {329}},\ \bibinfo {pages} {61}
  (\bibinfo {year} {2010})}\BibitemShut {NoStop}%
\bibitem [{\citenamefont {Bloembergen}\ and\ \citenamefont
  {Rowland}(1955)}]{Bloembergen1955}%
  \BibitemOpen
  \bibfield  {author} {\bibinfo {author} {\bibfnamefont {N.}~\bibnamefont
  {Bloembergen}}\ and\ \bibinfo {author} {\bibfnamefont {T.~J.}\ \bibnamefont
  {Rowland}},\ }\href {\doibase 10.1103/PhysRev.97.1679} {\bibfield  {journal}
  {\bibinfo  {journal} {Phys. Rev.}\ }\textbf {\bibinfo {volume} {97}},\
  \bibinfo {pages} {1679} (\bibinfo {year} {1955})}\BibitemShut {NoStop}%
\bibitem [{\citenamefont {{Lachman}}\ \emph {et~al.}(2015)\citenamefont
  {{Lachman}}, \citenamefont {{Young}}, \citenamefont {{Richardella}},
  \citenamefont {{Cuppens}}, \citenamefont {{HR}}, \citenamefont {{Anahory}},
  \citenamefont {{Meltzer}}, \citenamefont {{Kandala}}, \citenamefont
  {{Kempinger}}, \citenamefont {{Myasoedov}}, \citenamefont {{Huber}},
  \citenamefont {{Samarth}},\ and\ \citenamefont {{Zeldov}}}]{Lachman2015}%
  \BibitemOpen
  \bibfield  {author} {\bibinfo {author} {\bibfnamefont {E.}~\bibnamefont
  {{Lachman}}}, \bibinfo {author} {\bibfnamefont {A.~F.}\ \bibnamefont
  {{Young}}}, \bibinfo {author} {\bibfnamefont {A.}~\bibnamefont
  {{Richardella}}}, \bibinfo {author} {\bibfnamefont {J.}~\bibnamefont
  {{Cuppens}}}, \bibinfo {author} {\bibfnamefont {N.}~\bibnamefont {{HR}}},
  \bibinfo {author} {\bibfnamefont {Y.}~\bibnamefont {{Anahory}}}, \bibinfo
  {author} {\bibfnamefont {A.~Y.}\ \bibnamefont {{Meltzer}}}, \bibinfo {author}
  {\bibfnamefont {A.}~\bibnamefont {{Kandala}}}, \bibinfo {author}
  {\bibfnamefont {S.}~\bibnamefont {{Kempinger}}}, \bibinfo {author}
  {\bibfnamefont {Y.}~\bibnamefont {{Myasoedov}}}, \bibinfo {author}
  {\bibfnamefont {M.~E.}\ \bibnamefont {{Huber}}}, \bibinfo {author}
  {\bibfnamefont {N.}~\bibnamefont {{Samarth}}}, \ and\ \bibinfo {author}
  {\bibfnamefont {E.}~\bibnamefont {{Zeldov}}},\ }\href@noop {} {\bibfield
  {journal} {\bibinfo  {journal} {arXiv:1506.05114}\ } (\bibinfo {year}
  {2015})}\BibitemShut {NoStop}%
\bibitem [{\citenamefont {Lee}\ \emph {et~al.}(2015)\citenamefont {Lee},
  \citenamefont {Kim}, \citenamefont {Lee}, \citenamefont {Billinge},
  \citenamefont {Zhong}, \citenamefont {Schneeloch}, \citenamefont {Liu},
  \citenamefont {Valla}, \citenamefont {Tranquada}, \citenamefont {Gu},\ and\
  \citenamefont {Davis}}]{Lee2015}%
  \BibitemOpen
  \bibfield  {author} {\bibinfo {author} {\bibfnamefont {I.}~\bibnamefont
  {Lee}}, \bibinfo {author} {\bibfnamefont {C.~K.}\ \bibnamefont {Kim}},
  \bibinfo {author} {\bibfnamefont {J.}~\bibnamefont {Lee}}, \bibinfo {author}
  {\bibfnamefont {S.~J.~L.}\ \bibnamefont {Billinge}}, \bibinfo {author}
  {\bibfnamefont {R.}~\bibnamefont {Zhong}}, \bibinfo {author} {\bibfnamefont
  {J.~A.}\ \bibnamefont {Schneeloch}}, \bibinfo {author} {\bibfnamefont
  {T.}~\bibnamefont {Liu}}, \bibinfo {author} {\bibfnamefont {T.}~\bibnamefont
  {Valla}}, \bibinfo {author} {\bibfnamefont {J.~M.}\ \bibnamefont
  {Tranquada}}, \bibinfo {author} {\bibfnamefont {G.}~\bibnamefont {Gu}}, \
  and\ \bibinfo {author} {\bibfnamefont {J.~C.~S.}\ \bibnamefont {Davis}},\
  }\href {\doibase 10.1073/pnas.1424322112} {\bibfield  {journal} {\bibinfo
  {journal} {Proc. Na. Acad. Sci. U.S.A}\ }\textbf {\bibinfo {volume} {112}},\
  \bibinfo {pages} {1316} (\bibinfo {year} {2015})}\BibitemShut {NoStop}%
\bibitem [{\citenamefont {Li}\ \emph {et~al.}(2010)\citenamefont {Li},
  \citenamefont {Wang}, \citenamefont {Kan}, \citenamefont {Guo}, \citenamefont
  {He}, \citenamefont {Wang}, \citenamefont {Wang}, \citenamefont {Wong},
  \citenamefont {Wang},\ and\ \citenamefont {Xie}}]{Li2010}%
  \BibitemOpen
  \bibfield  {author} {\bibinfo {author} {\bibfnamefont {H.~D.}\ \bibnamefont
  {Li}}, \bibinfo {author} {\bibfnamefont {Z.~Y.}\ \bibnamefont {Wang}},
  \bibinfo {author} {\bibfnamefont {X.}~\bibnamefont {Kan}}, \bibinfo {author}
  {\bibfnamefont {X.}~\bibnamefont {Guo}}, \bibinfo {author} {\bibfnamefont
  {H.~T.}\ \bibnamefont {He}}, \bibinfo {author} {\bibfnamefont
  {Z.}~\bibnamefont {Wang}}, \bibinfo {author} {\bibfnamefont {J.~N.}\
  \bibnamefont {Wang}}, \bibinfo {author} {\bibfnamefont {T.~L.}\ \bibnamefont
  {Wong}}, \bibinfo {author} {\bibfnamefont {N.}~\bibnamefont {Wang}}, \ and\
  \bibinfo {author} {\bibfnamefont {M.~H.}\ \bibnamefont {Xie}},\ }\href
  {http://stacks.iop.org/1367-2630/12/i=10/a=103038} {\bibfield  {journal}
  {\bibinfo  {journal} {New Journal of Physics}\ }\textbf {\bibinfo {volume}
  {12}},\ \bibinfo {pages} {103038} (\bibinfo {year} {2010})}\BibitemShut
  {NoStop}%
\bibitem [{\citenamefont {Bansal}\ \emph {et~al.}(2011)\citenamefont {Bansal},
  \citenamefont {Kim}, \citenamefont {Edrey}, \citenamefont {Brahlek},
  \citenamefont {Horibe}, \citenamefont {Iida}, \citenamefont {Tanimura},
  \citenamefont {Li}, \citenamefont {Feng}, \citenamefont {Lee}, \citenamefont
  {Gustafsson}, \citenamefont {Andrei},\ and\ \citenamefont {Oh}}]{Bansal2011}%
  \BibitemOpen
  \bibfield  {author} {\bibinfo {author} {\bibfnamefont {N.}~\bibnamefont
  {Bansal}}, \bibinfo {author} {\bibfnamefont {Y.~S.}\ \bibnamefont {Kim}},
  \bibinfo {author} {\bibfnamefont {E.}~\bibnamefont {Edrey}}, \bibinfo
  {author} {\bibfnamefont {M.}~\bibnamefont {Brahlek}}, \bibinfo {author}
  {\bibfnamefont {Y.}~\bibnamefont {Horibe}}, \bibinfo {author} {\bibfnamefont
  {K.}~\bibnamefont {Iida}}, \bibinfo {author} {\bibfnamefont {M.}~\bibnamefont
  {Tanimura}}, \bibinfo {author} {\bibfnamefont {G.-H.}\ \bibnamefont {Li}},
  \bibinfo {author} {\bibfnamefont {T.}~\bibnamefont {Feng}}, \bibinfo {author}
  {\bibfnamefont {H.-D.}\ \bibnamefont {Lee}}, \bibinfo {author} {\bibfnamefont
  {T.}~\bibnamefont {Gustafsson}}, \bibinfo {author} {\bibfnamefont
  {E.}~\bibnamefont {Andrei}}, \ and\ \bibinfo {author} {\bibfnamefont
  {S.}~\bibnamefont {Oh}},\ }\href {\doibase
  http://dx.doi.org/10.1016/j.tsf.2011.07.033} {\bibfield  {journal} {\bibinfo
  {journal} {Thin Solid Films}\ }\textbf {\bibinfo {volume} {520}},\ \bibinfo
  {pages} {224} (\bibinfo {year} {2011})}\BibitemShut {NoStop}%
\bibitem [{\citenamefont {Wang}\ \emph {et~al.}(2011)\citenamefont {Wang},
  \citenamefont {Li}, \citenamefont {Guo}, \citenamefont {Ho},\ and\
  \citenamefont {Xie}}]{Wang2011}%
  \BibitemOpen
  \bibfield  {author} {\bibinfo {author} {\bibfnamefont {Z.~Y.}\ \bibnamefont
  {Wang}}, \bibinfo {author} {\bibfnamefont {H.~D.}\ \bibnamefont {Li}},
  \bibinfo {author} {\bibfnamefont {X.}~\bibnamefont {Guo}}, \bibinfo {author}
  {\bibfnamefont {W.~K.}\ \bibnamefont {Ho}}, \ and\ \bibinfo {author}
  {\bibfnamefont {M.~H.}\ \bibnamefont {Xie}},\ }\href {\doibase
  http://dx.doi.org/10.1016/j.jcrysgro.2011.08.029} {\bibfield  {journal}
  {\bibinfo  {journal} {Journal of Crystal Growth}\ }\textbf {\bibinfo {volume}
  {334}},\ \bibinfo {pages} {96} (\bibinfo {year} {2011})}\BibitemShut
  {NoStop}%
\bibitem [{\citenamefont {Tarakina}\ \emph {et~al.}(2012)\citenamefont
  {Tarakina}, \citenamefont {Schreyeck}, \citenamefont {Borzenko},
  \citenamefont {Schumacher}, \citenamefont {Karczewski}, \citenamefont
  {Brunner}, \citenamefont {Gould}, \citenamefont {Buhmann},\ and\
  \citenamefont {Molenkamp}}]{Tarakina2012}%
  \BibitemOpen
  \bibfield  {author} {\bibinfo {author} {\bibfnamefont {N.~V.}\ \bibnamefont
  {Tarakina}}, \bibinfo {author} {\bibfnamefont {S.}~\bibnamefont {Schreyeck}},
  \bibinfo {author} {\bibfnamefont {T.}~\bibnamefont {Borzenko}}, \bibinfo
  {author} {\bibfnamefont {C.}~\bibnamefont {Schumacher}}, \bibinfo {author}
  {\bibfnamefont {G.}~\bibnamefont {Karczewski}}, \bibinfo {author}
  {\bibfnamefont {K.}~\bibnamefont {Brunner}}, \bibinfo {author} {\bibfnamefont
  {C.}~\bibnamefont {Gould}}, \bibinfo {author} {\bibfnamefont
  {H.}~\bibnamefont {Buhmann}}, \ and\ \bibinfo {author} {\bibfnamefont
  {L.~W.}\ \bibnamefont {Molenkamp}},\ }\href {\doibase 10.1021/cg201636g}
  {\bibfield  {journal} {\bibinfo  {journal} {Crystal Growth \& Design}\
  }\textbf {\bibinfo {volume} {12}},\ \bibinfo {pages} {1913} (\bibinfo {year}
  {2012})}\BibitemShut {NoStop}%
\bibitem [{\citenamefont {Guo}\ \emph {et~al.}(2013)\citenamefont {Guo},
  \citenamefont {Xu}, \citenamefont {Liu}, \citenamefont {Zhao}, \citenamefont
  {Dai}, \citenamefont {He}, \citenamefont {Wang}, \citenamefont {Liu},
  \citenamefont {Ho},\ and\ \citenamefont {Xie}}]{Guo2013}%
  \BibitemOpen
  \bibfield  {author} {\bibinfo {author} {\bibfnamefont {X.}~\bibnamefont
  {Guo}}, \bibinfo {author} {\bibfnamefont {Z.~J.}\ \bibnamefont {Xu}},
  \bibinfo {author} {\bibfnamefont {H.~C.}\ \bibnamefont {Liu}}, \bibinfo
  {author} {\bibfnamefont {B.}~\bibnamefont {Zhao}}, \bibinfo {author}
  {\bibfnamefont {X.~Q.}\ \bibnamefont {Dai}}, \bibinfo {author} {\bibfnamefont
  {H.~T.}\ \bibnamefont {He}}, \bibinfo {author} {\bibfnamefont {J.~N.}\
  \bibnamefont {Wang}}, \bibinfo {author} {\bibfnamefont {H.~J.}\ \bibnamefont
  {Liu}}, \bibinfo {author} {\bibfnamefont {W.~K.}\ \bibnamefont {Ho}}, \ and\
  \bibinfo {author} {\bibfnamefont {M.~H.}\ \bibnamefont {Xie}},\ }\href
  {\doibase doi:http://dx.doi.org/10.1063/1.4802797} {\bibfield  {journal}
  {\bibinfo  {journal} {Applied Physics Letters}\ }\textbf {\bibinfo {volume}
  {102}},\ \bibinfo {pages} {151604} (\bibinfo {year} {2013})}\BibitemShut
  {NoStop}%
\bibitem [{\citenamefont {Schreyeck}\ \emph {et~al.}(2013)\citenamefont
  {Schreyeck}, \citenamefont {Tarakina}, \citenamefont {Karczewski},
  \citenamefont {Schumacher}, \citenamefont {Borzenko}, \citenamefont
  {Br\"une}, \citenamefont {Buhmann}, \citenamefont {Gould}, \citenamefont
  {Brunner},\ and\ \citenamefont {Molenkamp}}]{Schreyeck2013}%
  \BibitemOpen
  \bibfield  {author} {\bibinfo {author} {\bibfnamefont {S.}~\bibnamefont
  {Schreyeck}}, \bibinfo {author} {\bibfnamefont {N.~V.}\ \bibnamefont
  {Tarakina}}, \bibinfo {author} {\bibfnamefont {G.}~\bibnamefont
  {Karczewski}}, \bibinfo {author} {\bibfnamefont {C.}~\bibnamefont
  {Schumacher}}, \bibinfo {author} {\bibfnamefont {T.}~\bibnamefont
  {Borzenko}}, \bibinfo {author} {\bibfnamefont {C.}~\bibnamefont {Br\"une}},
  \bibinfo {author} {\bibfnamefont {H.}~\bibnamefont {Buhmann}}, \bibinfo
  {author} {\bibfnamefont {C.}~\bibnamefont {Gould}}, \bibinfo {author}
  {\bibfnamefont {K.}~\bibnamefont {Brunner}}, \ and\ \bibinfo {author}
  {\bibfnamefont {L.~W.}\ \bibnamefont {Molenkamp}},\ }\href {\doibase
  doi:http://dx.doi.org/10.1063/1.4789775} {\bibfield  {journal} {\bibinfo
  {journal} {Applied Physics Letters}\ }\textbf {\bibinfo {volume} {102}},\
  \bibinfo {pages} {041914} (\bibinfo {year} {2013})}\BibitemShut {NoStop}%
\bibitem [{\citenamefont {Tarakina}\ \emph {et~al.}(2014)\citenamefont
  {Tarakina}, \citenamefont {Schreyeck}, \citenamefont {Luysberg},
  \citenamefont {Grauer}, \citenamefont {Schumacher}, \citenamefont
  {Karczewski}, \citenamefont {Brunner}, \citenamefont {Gould}, \citenamefont
  {Buhmann}, \citenamefont {Dunin-Borkowski},\ and\ \citenamefont
  {Molenkamp}}]{Tarakina2014}%
  \BibitemOpen
  \bibfield  {author} {\bibinfo {author} {\bibfnamefont {N.~V.}\ \bibnamefont
  {Tarakina}}, \bibinfo {author} {\bibfnamefont {S.}~\bibnamefont {Schreyeck}},
  \bibinfo {author} {\bibfnamefont {M.}~\bibnamefont {Luysberg}}, \bibinfo
  {author} {\bibfnamefont {S.}~\bibnamefont {Grauer}}, \bibinfo {author}
  {\bibfnamefont {C.}~\bibnamefont {Schumacher}}, \bibinfo {author}
  {\bibfnamefont {G.}~\bibnamefont {Karczewski}}, \bibinfo {author}
  {\bibfnamefont {K.}~\bibnamefont {Brunner}}, \bibinfo {author} {\bibfnamefont
  {C.}~\bibnamefont {Gould}}, \bibinfo {author} {\bibfnamefont
  {H.}~\bibnamefont {Buhmann}}, \bibinfo {author} {\bibfnamefont {R.~E.}\
  \bibnamefont {Dunin-Borkowski}}, \ and\ \bibinfo {author} {\bibfnamefont
  {L.~W.}\ \bibnamefont {Molenkamp}},\ }\href {\doibase 10.1002/admi.201400134}
  {\bibfield  {journal} {\bibinfo  {journal} {Advanced Materials Interfaces}\
  }\textbf {\bibinfo {volume} {1}},\ \bibinfo {pages} {1400134} (\bibinfo
  {year} {2014})}\BibitemShut {NoStop}%
\bibitem [{\citenamefont {Wang}\ \emph {et~al.}(2014)\citenamefont {Wang},
  \citenamefont {Lian},\ and\ \citenamefont {Zhang}}]{Wang2014}%
  \BibitemOpen
  \bibfield  {author} {\bibinfo {author} {\bibfnamefont {J.}~\bibnamefont
  {Wang}}, \bibinfo {author} {\bibfnamefont {B.}~\bibnamefont {Lian}}, \ and\
  \bibinfo {author} {\bibfnamefont {S.-C.}\ \bibnamefont {Zhang}},\ }\href
  {http://link.aps.org/doi/10.1103/PhysRevB.89.085106} {\bibfield  {journal}
  {\bibinfo  {journal} {Physical Review B}\ }\textbf {\bibinfo {volume} {89}},\
  \bibinfo {pages} {085106} (\bibinfo {year} {2014})}\BibitemShut {NoStop}%
\end{thebibliography}%
\end{document}